\documentclass[12pt,a4paper]{article}
\usepackage{graphics}
\usepackage{amsfonts}
\usepackage{amstex}

\newcommand{\preprint}[1]{\begin{table}[t]  
           \begin{flushright}               
           \begin{large}{#1}\end{large}     
           \end{flushright}                 
           \end{table}}                     
\newcommand{\e}{\varepsilon}
\newcommand{\re} {\operatorname{Re}}
\newcommand{\im} {\operatorname{Im}}
\newcommand{\pa}{\partial}

\input{psfig}
\begin{document}
\begin{titlepage}

\preprint{hep-th/9809133\\TAUP-2518-98}

\vspace{2cm}

\begin{center}
{\bf\LARGE Precision ``Measurements" of the $Q \bar{Q}$
  Potential in MQCD
}\footnote{
Work supported in part by the US-Israel Binational Science
 Foundation,
by GIF - the German-Israeli Foundation for Scientific Research,
 and by
the Israel Science Foundation.
}\\
\vspace{1.5cm}
{\bf Y. Kinar  $\;\;$ E. Schreiber $\;\;$ J. Sonnenschein}\\
\vspace{.7cm}
{\em Raymond and Beverly Sackler Faculty of Exact Sciences\\
 School of Physics and Astronomy\\
 Tel Aviv University, Ramat Aviv, 69978, Israel\\
e-mail: \{yaronki,schreib,cobi\}\verb+@+post.tau.ac.il}
\end{center}
\vskip 0.61 cm

\begin{abstract}
We compute the M theory corrections to the confining linear potential between
a quark and an anti--quark in ${\cal N} = 1$ Super Yang--Mills  theory.
We find a constant term, 
and a term exponentially small with characteristic length of
$\Lambda_{QCD}^{-1}$. 

The potential in the MQCD setup that corresponds to softly broken ${\cal N} = 1$
SYM  is found to have a similar behavior. 
\end{abstract}

\end{titlepage}

\baselineskip 18pt

\section{Introduction}
During the last few years new light was shed on gauge theories via
world volume  field theories
 on D--branes stretched between two NS5 or D--branes, 
in IIA or IIB string theories
\cite{HaWi} (see also \cite{GiKu} and references therein).
 In particular, the brane configurations of 
$D = 3+1$, ${\cal N} = 1$ Super Yang--Mills (SYM) and Super QCD were analyzed.
The M theory versions of those theories, living on the M5 brane 
\cite{HOO,Wit2,BIKSY}, give a powerful new tool to deal with those
theories (It should be stressed, though, that the M theory versions
are not identical to the original theories, although they lie in the
same universality class). In particular, the
identification of the QCD string as an M2 membrane ending on the
M5 \cite{Wit2},
enables the calculation of the quark anti--quark potential via the
MQCD string tension \cite{HSZ}. The leading order of the potential was
 found to be linear in the separation $l$ between the quark anti--quark pair, 
$E \propto l$.
This result  which is valid, of course, only for large separations, 
(for $l \gg \Lambda_{QCD}^{-1}$  where the
meson is ``stretched''), is an indication of  confinement.

In this note we  ``measure" the potential with higher precision than
that of \cite{HSZ}. 
We derive the Nambu--Goto action that corresponds to an $SU(N)$ gauge theories.
 In the case of 
$SU(N = 2)$  we explicitly determine the sub--leading terms of
the potential  dependence on $l$, in  the limit $\Lambda_{QCD} \ll 1$ (in M
theory units). 
We find that the corrections to the linear term include a constant term, 
and a term exponentially small with characteristic length of
$\Lambda_{QCD}^{-1}$. 
We must remark, though, that because the MQCD string is not a BPS
state, its tension is not protected  when we vary the
parameters of the theory.

Similar nature of the sub--leading corrections to the potential are also
detected in calculations involving MQCD setups that correspond to 
softly broken ${\cal N}=1$ SYM theory. 

Our work involves calculations similar to ones performed recently in the
context of the
AdS/CFT correspondence \cite{Mal}--\cite{Min}.
It turns out that the  next to leading corrections found in our analysis 
are of the same nature of those found in a supergravity setup that corresponds
to the pure four dimensional YM theory \cite{BISY2,GrOl}. 
We also find no trace of a $\propto 1/l$
 L\"{u}scher \cite{Luscher} term correction which 
 is believed to exist in lattice simulations \cite{Tep} of YM theory. 
We argue that incorporating quantum fluctuations  of the string world-sheet 
may induce such a term as was also the case in ref. \cite{Luscher}.

\section{$Q\bar{Q}$ potential in ${\cal N} = 1$ MQCD}

The potential energy of a quark anti-quark pair translates in the MQCD
language to the area of the M-theory membrane 
that corresponds to the QCD string. Therefore, the extraction of the 
potential energy involves the construction of the MQCD setup \cite{HSZ},
introducing a convenient parameterization of  the membrane, and computing
its area.
We begin by reviewing the MQCD setup of  \cite{HSZ}.

\subsection{The setup}

 The MQCD description of ${\cal N} = 1$ SQCD is 
given in terms of a five-brane (M5)
  embedded in
$R^{10} \times S^1 \equiv R \times R^4 \times Y$, 
where the 11-th compact dimension
$S^1$ is a circle of radius $R$.
The M5  has the form $R^4 \times \Sigma$, where $\Sigma$
is a Riemann surface in the space $Y$ which is a three dimensional 
complex space with coordinates
$v = x_4 + i x_5 \,,\, w=x_7+ i x_8 \,,\,  t = e^{-s/R}$ where 
$s = x_6 + i R x_{10} $~. The setup does not extend in the additional 
non compact dimension, and we can express this as the requirement $x_9 = 0$.
  
\begin{figure}[h!]
\begin{center}
\resizebox{0.6\textwidth}{!}{\includegraphics{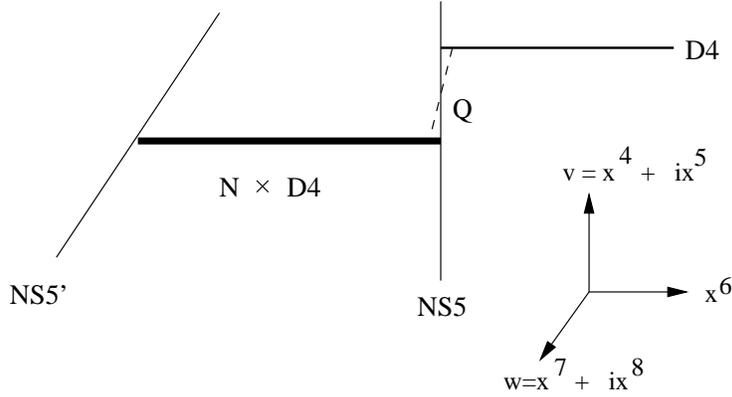}}
\end{center}
\caption{The ${\cal N}=1$ branes configuration}
\label{fig:branes}
\end{figure}

In the limit $R \rightarrow 0$ the M5 reduces to 
the well known type IIA brane configuration of figure (\ref{fig:branes})
\cite{EGK}. In this configuration the flavor degrees of freedom are associated
 with the semi-infinite
D4 branes stretched to the right of the NS5 brane. 
An alternative setup includes  semi-infinite
D4 branes stretched both to the right of the NS5 brane and to the left of the NS5' brane. 
This later configuration is considered in section 2.4.
Although we have quarks in our setup they are very heavy and we take 
for $\Sigma$ the Super Yang--Mills $SU(N)$ curve \cite{HOO,Wit2,BIKSY} 
\begin{eqnarray}
v \cdot w = \zeta \\
v^N = \zeta^{N/2} t 
\end{eqnarray}
where $\zeta$ is a complex constant. We measure distances in M-theory
units, namely, in units where the Planck length is set to be $l_p=1$.
We see that for each $t$ there
are $N$ distinct $(v,w)$ points, differing by powers of the phase 
 $e^{2 \pi i / N}$, and therefore there are $N$ branches of the M5 brane. 

The MQCD M5 is characterized by two 
dimensionful parameters, $R$ and $\zeta$, whereas the ${\cal N}=1$ SYM 
has only a single scale $\Lambda_{QCD}$. Matching  the string and 
domain wall tensions of MQCD with the corresponding ones of the large $N$
SYM requires the following assignments
 \cite{Wit2}:
\begin{eqnarray}
\label{RofL}
R     & \sim & {1 \over {N \Lambda_{QCD}}} \\
\label{zofL}
|\zeta| & \sim & N^2 \Lambda_{QCD}^4
\end{eqnarray}
The dependence of $R$ and $\zeta$ upon $\Lambda_{QCD}$ is 
dictated from  dimensional analysis. 
As we shall deal with the small $N$ regime, and are not interested 
in numerical factors in the $\Lambda_{QCD}$ language, the $N$ dependence
of (\ref{RofL}),(\ref{zofL}) will not be used by us.  

Quarks and W--bosons are described in MQCD by membranes wrapped around
the 11-th dimension. This leads to open strings of type IIA
in the limit $R \rightarrow 0$. The boundaries of such membranes must be
curves on $\Sigma$ with zero winding-number around the 11-th
dimension \cite {Mik}. Such membranes may have the topology of a disk
or of a cylinder. It was shown \cite {Mik,HeYi} that membranes with the
cylinder topology correspond to vector supermultyplets, while those
with the disk topology correspond to chiral ones. In \cite{HSZ,EPR} it
was demonstrated that in our setup we cannot describe a membrane
representing a single quark, hence quarks are confined.

\begin{figure}[h!]
\begin{center}
\resizebox{0.4\textwidth}{!}{\includegraphics{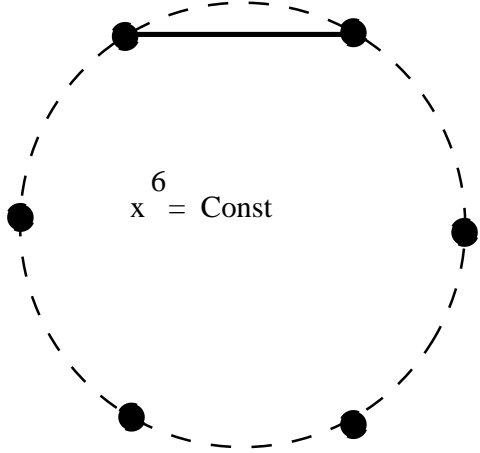}}
\end{center}
\caption{The QCD string for N=6. this is the $v$ plane for a constant $x_6$ }
\label{fig:string}
\end{figure}

The construction we are interested in is of a ``meson'', that is, 
of a quark and an anti--quark connected by a QCD string. The membrane 
corresponding to the QCD string is stretching in one direction along
$R^4$, while in the other direction it interpolates between two adjacent 
branches of the M5 brane (see figure (\ref{fig:string})). 
This membrane will not vary in the 
$x_{10}$ direction, so we might take
$x_{10} = 0$, and have $s = x_6$. The length of the interpolating segment for a
certain $s$, which is the width of the membrane, is (as we will see shortly) \cite {HSZ}
\begin{equation}
\label{width}
2 \sqrt{2 \zeta} \sin(\pi/N)\sqrt{ \cosh(2s/RN)}
\end{equation}
(From here on $\zeta$ stands for $|\zeta|$). 
The membrane corresponding to a fundamental type IIA string (``quark'')  transforms {\em smoothly} into a QCD string and back into a fundamental
string.
The membrane's area is the total energy of the ``meson''
in M-theory units. This area tends  to infinity as the 
quarks' masses  approach  infinity, so we need to subtract those
masses. We consider only the part of the membrane that
 behaves as a 'QCD string membrane' rather than 'quark membrane' 
(fundamental string in the type IIA string theory). We first characterize the point of the transition -- this is the point where the tension of the QCD string is the same as the type IIA string tension:
\begin{equation}
\label{sametension}
2 \sqrt{2 \zeta} \sqrt {\cosh (2s/RN)} = 2 \pi R
\end{equation}

Once we make all our calculations up to this value of $s$, the quarks, 
whose masses we wish to subtract, are simply QCD strings with 
constant coordinates in $R^4$ and $s$, going from the maximal value we found 
(which we will eventually take to infinity) to zero.

\subsection{The area of the membrane}

The QCD string is associated with  an M2  stretched in the $Y$ space, as well 
as in one spatial dimension of $R^4$, that we  take as $x \equiv x_3$.
For each $(x,s)$, the membrane's slice is a straight line segment in 
$(v,w)$ with both ends on $\Sigma$. 
We have the freedom to rotate the $v$ plane so that this segment will vary
only in $x_5$, and likewise for $w$ and $x_8$. Therefore, we can write 
a parameterization of the membrane as 
\footnote{Actually, in this parameterization, $v^N = -\zeta^{N/2} e^{-s/R}$,
so it corresponds to $x_{10} = \pi$. This, however, has no physical 
significance, and we shall continue to treat $s$ and $x_6$ interchangeably.}
\begin{eqnarray}
\label{memparam}
\vec{P}(x,\rho) & = & (x_3\,,\,x_4\,,\,x_5\,,\,x_6\,,\,x_7\,,\,x_8) \nonumber \\
       & = & (x\,,\,\sqrt{\zeta} e^ {-{s(x) / {R N}}} \cos (\pi/N) 
             \,,\, \rho \sqrt{\zeta} e^ {-{s(x) / {R N}}} \sin(\pi/N) \,,\, \\
       &   &  s(x)\,,\, \sqrt{\zeta} e^ {  s(x) / {R N} } \cos (\pi/N) 
             \,,\, -\rho \sqrt{\zeta} e^ {s(x) / {R N}} \sin(\pi/N) ) \nonumber
\end{eqnarray}
with $-1 \le \rho \le 1$. For $\rho = \pm 1$, $\vec{P}$ is on $\Sigma$ and the distance between these two points gives the string's width (\ref{width}).

The function $s(x) $ is determined by minimizing the area of the
membrane. The width (in the $(v,w)$ space) of the QCD string 
decreases as $s$ approaches $0$, so the  membrane has an
``incentive'' to achieve small values of $s$. On the other hand, of
course, this involves a large detour in $s$. The best compromise
is achieved, as we shall see, for a minimal value of $s$, being 
$s_0 \equiv s(0)$, exponentially small in $l$.

Taking the membrane's area as an action, we can write a Lagrangian
which is an area element in the form 
\begin{equation}
{\cal L} = \left| \left({{\partial \vec{P}} \over {\partial \rho}}\right) 
           \wedge 
           \left({{\partial \vec{P}} \over {\partial x   }} \right) \right|
=\sqrt{det[h_{\alpha\beta}]}
\end{equation}
where $h_{\alpha\beta}=\pa_\alpha \vec P \cdot \pa_\beta \vec P$,
and $\alpha$ and $\beta$ correspond to $\rho$ and $x$.
Inserting the explicit expression of (\ref{memparam}) one finds  
\begin{equation}
{\cal L} = \sqrt {2 \zeta} \sin(\pi/N) \sqrt{b + a \rho^2}
\end{equation}
where
\begin{eqnarray}
a & \equiv & {{2 \zeta} \over {R^2 N^2}} \sin^2 (\pi/N) {s'}^2\\
b & \equiv & \cosh ({{2 s} / {R N}}) \left(1 + {s'}^2 + {{2 \zeta} \over
    {R^2 N^2}} \cos^2 (\pi/N)  \cosh ({{2 s} / {R N}}) {s'}^2\right)
\end{eqnarray} 

integrating over $ \rho $ we get : 
\begin{equation}
{\cal L} = \sqrt {2 \zeta} \sin {(\pi/N)} \left( \sqrt {a+b} + {b
    \over {2 \sqrt{a}}} \left(\log (\sqrt{a+b}\,+\sqrt{a}) \: - \: \log
    (\sqrt{a+b} \,- \sqrt{a}) \right) \right)
\end{equation} 

We look at the limit $\Lambda_{QCD} \ll 1 $ (in M-theory units
\footnote{in those units, the eleven-dimensional Planck length is set
  to one, $l_{p} = 1$. The dimensionful condition is $\Lambda_{QCD}
  \ll l_p^{-1}$.}), so by (\ref{RofL},\ref{zofL}), for not too big an $N$,
\begin{equation}
\label{lambdalimit}
{\zeta \over {R^2 N^2}} \ll 1
\end{equation}

and therefore:
\begin{equation}
{\cal L} = 2 \sqrt {2 \zeta} \sin(\pi/N) \sqrt{\cosh ({{2 s}/{R N}})}
\sqrt {1 + {s'}^2 + {{2\zeta} \over {R^2 N^2}} \cos^2(\pi/N) 
\cosh ({{2 s}/{R N}}) {s'}^2}
\end{equation}

which has a very simple form for $SU(2)$: 
\begin{equation}
{\cal L} = 2 \sqrt {2 \zeta} \sqrt {\cosh ({s/R})} \sqrt {1+{s'}^2}
\end{equation}

This form has a simple interpretation. It corresponds to the area 
 ignoring the change in the $(v,w)$ 
coordinates of $\vec{P}(x,\rho = \pm 1)$, as a function of $x$, and retaining
only its $(x,s)$ change.
$\sqrt {1+{s'}^2}$ is then the length of the line element of
the QCD string in the $(x,s)$ plane, and 
$2 \sqrt {2 \zeta} \sqrt {\cosh ({s/R})}$ is  the width of
the membrane in the transverse $(v,w)$ plane. This width, then, provides
naturally the metric, which is scalar diagonal. 

However, although the change in $x_4$ and $x_7$ is suppressed by the factor 
(\ref{lambdalimit}), we cannot ignore it if it tends to infinity with $s$.
The only case in which we can ignore this change is when 
$x_4 \equiv x_7 \equiv 0$, that is, the case $N=2$. 
Hence, this interpretation is valid only for $N = 2$,
where, indeed, $x_4 \equiv x_7 \equiv 0$.  

The Hamiltonian of the $N=2$ action is 
\begin{equation}
{\cal H} = {{-2 \sqrt {2 \zeta} \sqrt {\cosh ({s/R})}} \over \sqrt {1+{s'}^2}} 
\end {equation}

We shall first treat the case in which the quark and the anti--quark are on
the same NS5 brane (in the IIA picture), that is, they both reside in the
$s \rightarrow +\infty$ limit. Due to the symmetry of the problem we have 
$ s'(0) = 0 $, and
defining $ s_0 \equiv s(0) $ we can write (as the Lagrangian
does not depend explicitly on $ x $): 
\begin{equation}
{\cal H}(s,s') ={{-2 \sqrt {2 \zeta} \sqrt {\cosh ({s/R})}} \over
         \sqrt {1+{s'}^2}}  = -2 \sqrt {2 \zeta} \sqrt {\cosh ({s_0/R})}
         = {\cal H}(s_0,0)
\end{equation}

This differential equation can be written as: 
\begin{equation}
s'(x) = \pm \sqrt {\cosh(s/R)/\cosh(s_0/R) \,-\, 1}
\end{equation}

which gives : 
\begin{equation}
x = \pm \int_{s_0}^{s(x)} {ds \over \sqrt {\cosh (s/R)/\cosh (s_0/R) \,-\, 1}}
\end{equation}

In particular, since by (\ref{sametension},\ref{lambdalimit}) we have 
$ s(l/2) \sim \log (R/\sqrt{\zeta}) \gg 1$, we can write:
\begin{equation}
{1\over 2}l = \int_{s_0}^\infty {{\sqrt{\cosh({s_0/R})} \,ds} \over
  \sqrt {{\cosh ({s/R}) - {\cosh({s_0/R})}}}}
\end{equation}

Inserting the solution we found into the action, we see that the energy of
the QCD string is:
\begin{eqnarray}
{1 \over 2}E' & = & \int_0^l {\cal L} \, dx \nonumber\\
              & = & \int_{s_0}^{s(l/2)} {\cal L} {dx \over ds} ds \nonumber \\
              & = & 2 \sqrt{2 \zeta} \int_{s_0}^\infty {{\cosh ({s/R}) ds} \over
    \sqrt {\cosh ({s/R}) - \cosh ({s_0/R})}} \\
             & = & 2 \sqrt{2 \zeta} \sqrt{\cosh({s_0/R})} ({1\over 2}l) + 
2 \sqrt{2 \zeta} \int_{s_0}^\infty  \sqrt {\cosh ({s/R}) - \cosh ({s_0/R})} \, ds \nonumber
\end{eqnarray}

from which we must subtract the (infinite) mass of the quark :
\begin{equation}
M_q = 2 \sqrt{2 \zeta} \int_0^\infty \sqrt{\cosh({s/R})} \: ds
\end{equation}

changing the integration variable to $ y = \cosh ({s/R}) - \cosh
({s_0/R}) $ and $z = \cosh ({s/R})$, and defining 
$ \e \equiv \cosh ({s_0/R}) - 1 $, we finally get: 
\begin {eqnarray}
\label{lofe}
l & = & 2R\sqrt{(1+\e)} \int_0^\infty {dy \over \sqrt{y(y+\e)(y+2+\e)}} \\
\label{eofe}
E & = & 2 \sqrt {2 \zeta} \sqrt{1+\e} \:l + 4 \sqrt 
{2 \zeta} R \: J(\e) - 4 \sqrt {2 \zeta} R \int_1^{1+\e} {{dz \sqrt{z}} \over \sqrt{z^2 - 1}}
\end{eqnarray}

where 
\begin{equation}
\label{jofe}
J(\e) \equiv \int_0^\infty dy {{\sqrt{y} - \sqrt{y+1+\e}} \over \sqrt{(y+\e)(y+2+\e)}}
\end{equation}

\begin{figure}[h!]
\begin{center}
\resizebox{0.4\textwidth}{!}{\includegraphics{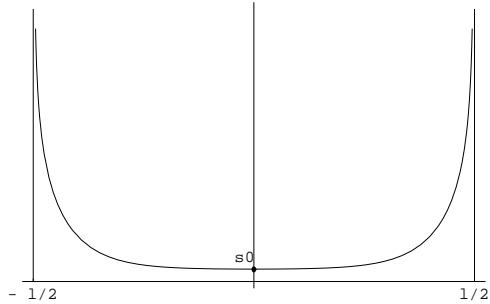}}
\end{center}
\caption{A characteristic shape of $s(x)$. The asymptotes represent ``bare'' quarks}
\label{fig:graph}
\end{figure}

\subsection{The potential dependence on the $Q\bar{Q}$ separation}

The distance $l$ as a function of $\e$ is shown in the appendix to take the 
following form 
\begin{equation}
\label{lofe2}
l = \left(-\sqrt{2} \log \e + 5 \sqrt{2} \log 2 - {3 \over {4 \sqrt{2}}} \e \log \e + O(\e) \right) R
\end{equation}
The explicit dependence of the energy $E$ on $\e$,  which is also 
computed in the appendix, reads
\begin{equation}
\label{Eofe2} 
E = 2 \sqrt{2 \zeta} \: l \left( 1 + {1 \over 2} \e - {1 \over 8} \e^2 + O(\e^3) + 2 (R/l) (- \kappa + {1 \over {2 \sqrt{2}}} \: \e \, \log {\e \over 32} - {1 \over {32 \sqrt{2}}} \: \e^2 \, \log {\e \over 32} + O(\e^2)) \right)
\end{equation}
with $\kappa = -J(0)$.

Inserting the value of $\e$  from the inverse of (\ref{lofe2}),
 we finally find the following   expression for $E$ as a function of $l$ 
\begin{equation}
E = 2 \sqrt{2 \zeta} \: l \left( 1 -2 (R/l) \kappa + 96 e ^{-2l / \sqrt{2}R} + O((R/l) e ^{-2l / \sqrt{2}R}) \right)
\end{equation}

It is easy to see how the corrections to the linear potential depend
on $\Lambda_{QCD}$, the parameter of ${\cal N}=1$ SYM theory, by translating 
the MQCD parameters according to (\ref{RofL},\ref{zofL}).
As a function of $l$, one correction is constant, 
while the other is essentially exponential in 
$-l/R \sim -\Lambda_{QCD} \: l$.
In fact, given the dependence on $l$, the dependence on $\Lambda_{QCD}$ 
is determined, since  $\Lambda_{QCD}$ is the only energy scale of 
the original theory.

Note that the potential does not include a L\"{u}scher term, namely, a term
of the form $-c/l$, where $c$ is a positive universal constant. 
Such a term is missing also in the Wilson line derived from the  
Ads/CFT correspondence \cite{GrOl}. Since the traces of such a term
 were observed in lattice simulations \cite{Tep} (and in phenomenological fits
to the heavy mesons spectra), the authors of ref. \cite{GrOl}
suggested that this may indicate that there is a phase transition.
The proposed scenario is that the L\"{u}scher term exists in the weak coupling
phase but is missing in the strong coupling phase.
It seems to us that the origin  of the discrepancy  is different and, in fact, 
is hidden in the work that led to the discovery of the $c/l$ term \cite{Luscher}.
The source of this term is the contribution of the Gaussian fluctuations 
around the classical solution. In the present work, as well as in those that 
follow the Ads/CFT duality, the computation was based solely on the 
classical solution and the quantum fluctuations have not been incorporated.
It is plausible   that the evaluation of the fluctuations around the classical 
configuration  in the stringy calculations of both MQCD and the AdS/CFT
duality  will produce a  term of the form $c/l$.

\subsection{$Q \bar{Q}$ potential of the symmetric MQCD setup}

We proceed now to discuss the case in which the quark and the anti--quark
are, asymptotically, at $s = \pm \infty$. In the IIA picture, they are on
the two NS5 branes, as opposed to the setup discussed so far,  in which they
are on the same NS5 brane.

The previous case gave rise to an even function $s(x)$, whereas now we
get an odd function. We expect the energy for this case
to be higher than the previous one. The reason is that if we reflect,
for negative $x$, the graph of the odd $s(x)$, we get an even membrane with the
same area. The solution found in the previous sections for the even case has,
of course, a smaller area. (The reflected membrane would not be in 
equilibrium for $x = 0 , s(x) = 0$).

The Lagrangian and Hamiltonian are equal to those of the previous
case, but, as $s(0) = 0$, the solution is now specified by the value of 
$r_0 \equiv s'(0)$, and
\begin{equation}
{\cal H}(s,s') ={-2{\sqrt {2 \zeta} \sqrt {\cosh ({s/R})}} \over
         \sqrt {1+{s'}^2}}  = -2 \sqrt {2 \zeta} {1 \over \sqrt{1+r_0^2}} 
         = {\cal H}(0,r_0)
\end{equation}

If we define for this case $\e = r_0^2$ and substitute 
$y = \cosh (s/R) - 1$, we get
\begin{equation}
 l = 2 R \int_0^\infty {dy \over \sqrt{y(y+2)(y(1+\e)+\e)}}
\end{equation}
and the energy,  after subtracting the quark mass, is 
\begin{equation}
 E = 2\sqrt{2\zeta} \: l \left({1 \over \sqrt{1+\e}} + 2 (R/l) \int_0^\infty dy {{\sqrt{y + \e/(1+\e)} - \sqrt {y+1}} \over {\sqrt{y(y+2)}}} \right)
\end{equation}

Using methods similar to those used in the even case, we get for the energy 
a stronger dependence on $\e$. This time we have 
\begin{equation}
E = 2 \sqrt{2 \zeta} \: l \left(1 - {1 \over 2} \e + O(\e^2) + 2 (R/l) (-\kappa - {1 \over {2 \sqrt{2}}} \: \e \, \log {\e \over 32} + {1 \over {2 \sqrt{2}}} \e + O(\e^2 \, \log \e))\right)
\end{equation}

We can, therefore, write $l$ only to the first order in $\e$. This happens to be identical to the even case, so we have for $\e$  
\begin{equation} 
\e = 32 e^{-l/\sqrt{2}R} + O(e^{-2l/\sqrt{2}R})
\end{equation}

Inserting this into the expression of the energy, we have for this case: 
\begin{equation}
E = 2 \sqrt{2 \zeta} \: l \left(1 - 2 (R/l) \kappa +2 (R/l) {16 \over {\sqrt{2}}} e^{-l/\sqrt{2}R} + O(e^{-2l/\sqrt{2}R}) \right)
\end{equation}

This potential has the same type of behavior as the one  with the asymmetric MQCD setup.
The corrections to the string tension term are a constant and an exponential term. However, it is clear that for small enough $\e$ (That is, large enough $l$)
this result is larger than the one we got for the even $s(x)$ case. 

\section{ $Q\bar{Q}$ potential in softly broken  ${\cal N} = 1$ MQCD}
We proceed now to analyze the  $Q\bar{Q}$ potential in a 
MQCD setup that corresponds to  softly broken  
 ${\cal N}=1 $ SYM.
In the field theory picture this is achieved by giving mass 
to the gaugino and in the context of brane configurations this is done 
by rotating one of the NS branes in a non-holomorphic way 
(e.g. rotation in the $(x_4,x_7)$ plane). 
The 
notations that we are using are those 
of \cite {Wit2} where  the brane is parameterized by two 
four dimensional complex null vectors $\vec p$ and $\vec q$,
\begin{equation}
{\vec p}^{\,\: 2} = {\vec q}^{\,\: 2} = 0 
\end{equation}
in the following form 
\begin{eqnarray}
(x_4,x_5,x_7,x_8) & = & \re{(\vec{p} \lambda + \vec{q} \lambda^{-1})} \\
x_6 & = & - R N c \re \log \lambda \\
x_{10} & = & -N \im \log \lambda
\end{eqnarray}

where $\lambda$ is a complex number and $c$ is a real number given by 
\begin{equation}
\label{cofpq}
- {\vec p} \cdot {\vec q} + {{R^2 N^2} \over 2} (1-c^2) = 0 
\end{equation}
In this notation we choose the ${\cal N}=1$ curve  to be  
\begin{equation}
{\vec p} = \sqrt{\zeta} (1,-i,0,0) \,\,\,\,\,\,\,\,  {\vec q} = \sqrt{\zeta} (0,0,1,-i)
\end{equation}

Using the rotation suggested in \cite {Evans,BaPa} we keep $\vec p$ the 
same as for ${\cal N}=1$ and choose 
\begin{equation}
\label{mishq}
{\vec q} = \sqrt{\zeta} (\sin \alpha, i \sin \alpha, \cos \alpha, -i \cos \alpha) 
\end{equation}

 For simplicity, we shall concentrate now on $SU(2)$ YM,
namely, considering the brane at a constant 
$s \equiv e^{-x_6}$ (that is, $\lambda = r e^{i \theta} \equiv { s^{1/ 2 R c}} \:e^{i \theta}$) 
we find the cross-section with the $v$ plane to be a circle of 
radius $\sqrt{\zeta}\,(r \, + \, {\sin \alpha \, / \, r})$,  
and the cross section 
with the $w$ plane a circle of radius $\sqrt{\zeta}\,{{\cos \alpha} \, /\, r}$.
 The distance between two points of equal $x_{10}$ is therefore  
\begin{equation}
2 \sqrt{2 \zeta} \sqrt{\cosh({s/Rc}) + \sin{\alpha}} = 2 \sqrt{2 \zeta \cosh({s/Rc}) + {\vec p} \cdot {\vec q}}
\end{equation}

Using the arguments given above for the supersymmetric $SU(2)$ theory we can 
write the Lagrangian of this theory, defining $\mu = \sin \alpha$, as 
\begin{equation}
{\cal L} = 2 \sqrt {2 \zeta} \sqrt {\cosh ({s/Rc}) + \mu} \sqrt {1+{s'}^2}
\end{equation}

and the expressions for the length and the energy of the QCD string  
\begin {eqnarray}
l & = & 2R \sqrt{(1+\mu + \e)} \int_0^\infty {dy \over \sqrt{y(y+\e)(y+2+\e)}} \\
E & = & 2 \sqrt {2 \zeta} \sqrt{1+\mu + \e} \:l + 4 \sqrt 
{2 \zeta} R \: J(\e) - 4 \sqrt {2 \zeta} R \int_1^{1+\e} {{dz \sqrt{z+ \mu}} \over \sqrt{z^2 - 1}}
\end{eqnarray}

where 
\begin{equation}
J(\e) \equiv \int_0^\infty dy {{\sqrt{y} - \sqrt{y+1+\mu+\e}} \over \sqrt{(y+\e)(y+2+\e)}}
\end{equation}

and we used (according to (\ref{lambdalimit})) : 
\begin{equation}
c = \sqrt{1 - {{\vec p} \cdot {\vec q} / 2 R^2}} = \sqrt{1 - {\mu \zeta / R^2}} \simeq 1
\end{equation}

Using the same methods as in the supersymmetric case we find 
\begin{eqnarray}
\log (\e / 32) & = & - {l \over {\sqrt{2} R}}{1 \over  \sqrt{1+\mu}} \left( 1 - 32({1 \over {2 \sqrt{1+\mu}}} - {1 \over 8}) e^{-l/(\sqrt{2}R\sqrt{1+\mu})}\right) + O(e^{-l/(\sqrt{2}R\sqrt{1+\mu})}) \nonumber \\
\e / 32 & = & e^{-l/(\sqrt{2}R\sqrt{1+\mu})} + {32 \over \sqrt{2}} {l \over R}{1 \over \sqrt{1+\mu}} ({1 \over {2 \sqrt{1+\mu}}} - {1 \over 8})  e^{-2l/(\sqrt{2}R\sqrt{1+\mu})} \\
        &   &  \mbox{} + O(e^{-2l/(\sqrt{2}R\sqrt{1+\mu})}) \nonumber
\end{eqnarray}
and for $E(l(\e),\e,\mu)$
\begin{eqnarray} 
E & = & 2 \sqrt{2 \zeta} \: l \left( \sqrt{1+\mu} + {\e \over {2 \sqrt{1+\mu}}} - {\e^2 \over {8 (1+\mu)^{3/2}}} + 
2 {R \over l} (- \kappa_\mu + {\e \over {2 \sqrt{2}}}  \, \log {\e \over 32} - {\e^2 \over {32 \sqrt{2}}} \, \log {\e \over 32}) \right) \nonumber\\
  &  & \mbox{} + O(R \e^2) + O(l \e^3)
\end{eqnarray}
where $-\kappa_\mu$ is the zeroth order term of $J$. 
 We finally have for the energy in the softly broken SYM theory
\begin{eqnarray}
E & = & 2 \sqrt{2 \zeta} \: l \left( \sqrt{1+\mu} -2 (R/l) \kappa_\mu +
  \frac{32}{(1+\mu)^{3/2}}
  (-5-\mu+8\sqrt{1+\mu})e^{-2l/(\sqrt{2}R\sqrt{1+\mu})} \right. \nonumber \\
  & & \left. \mbox{} + O((R/l) e ^{-2l / (\sqrt{2}R\sqrt{1+\mu})}) \right)
\end{eqnarray}
Using the relation of the rotation parameter $\alpha$ to the gaugino mass given in 
\cite{Evans,BaPa} we get in our notations the plausible result  
$m_\lambda = \tan{\alpha}$, and the pure YM theory is given by $\alpha = \pi / 2$ and $\mu=1$ .
In this  case the  energy  takes the form  
\begin{equation}
\label{Enons}
E = 2 \sqrt{2 \zeta} \: l \left( \sqrt{2} -2 (R/l) \kappa_1 + 16(8-3\sqrt{2})e^{-l/R} + O((R/l) e^{- l/R} ) \right)
\end{equation}
Comparing this result to the one of the ${\cal N}=1$ case  we see that it has 
the same principal functional
dependence on $l$. The string tension is modified, 
however, and so is the exponential correction, since the screening length 
is $R$ rather than $R/\sqrt{2}$ as in the supersymmetric case. 
As the equations (\ref{RofL},\ref{zofL}) give the relation between 
$R , \zeta$ and $\Lambda_{QCD}$ only upto dimensionless constants, the
significance of those modifications in field theory is unclear.
 
An interesting special case of this calculation is when $\mu=-1$,  
when the curve spans only the $x_6$ and $v$ coordinates and is singular 
at $|\lambda|=1$ (or $x_6=0$), where the corresponding QCD string's width 
is zero. 
The behavior we find for this case is very similar to the one described in \cite{BISY1} 
(${\cal N}=4$ at finite temperature).
 Once we regulate the integrals (by taking the upper limit as $s_{max}$ instead of $\infty$) 
we find that for $l \gg \sqrt{s_{max} R}$  the quarks are free 
(the favorite configuration is of two ''bare'' quarks with the QCD string located at $x_6$=0) 
while for $l \ll \sqrt{s_{max} R}$ we still have linear confinement as the leading order behavior. 

\section{Summary and Discussion}

Quark anti-quark potentials have been extracted recently from computations
of Wilson loops in various supergravity models which are conjectured duals of 
supersymmetric gauge theories in the large $N$ limit 
\cite{Mal}--\cite{Min}.
It is thus interesting to compare the calculations in the 
MQCD framework to those of the gauge/gravity dual models.
In both cases, in spite of the fact that the starting points 
seem very different,  
the calculations reduce eventually to computing the area of a
bosonic string world-sheet via  a Nambu--Goto action. 
Moreover, using a gauge $\sigma =x$ where  
$\sigma$ is a world-sheet coordinate and $x$ is a coordinate along which
the quark anti-quark pair is situated,  the MQCD and gravity calculations are
special cases of an integral of the following form \cite{KSS}
\begin{equation}
S=\int dx\sqrt{ f^2(s(x)) + g^2(s(x)) (\pa_x s)^2}
\end{equation}
where $s$ is a coordinate external to the $R^{1,3}$ space-time,
$s=x_6$ in the MQCD case and the fifth coordinate of the $AdS_5$ space
(or of its relatives).
 In fact, in the approach of Polyakov \cite{Polyakov}, $s$
is associated with the Liouville coordinate. It is straightforward to 
realize that $f^2(s),g^2(s) $ are both  $ \cosh(s/R)$ in $SU(2)$ MQCD,
 $(s/R_{Ads})^4,1$ in the Ads/CFT case \cite{Mal},
 $(s/R_{Ads})^4(1-(s_T/s)^4),1$ in finite temperature \cite{BISY1} and
 $(s/R_{Ads})^4,(1-(s_T/s)^4)^{-1}$ for the ``pure YM'' 
4 dimensional model  \cite{BISY2}.
Naturally we would like to compare the quark potentials  extracted from the
stringy computations. In particular the MQCD result for broken supersymmetry
(\ref{Enons}) can be compared with  the precise computation  of pure YM theory 
in four dimensions (eqn. (22) of \cite{GrOl}).
In both cases the corrections to the linear potential do not include a $c/l$
L\"uscher term, but only an exponential term of the form $le^{-bl}$, where  $b$
is proportional to $\Lambda_{QCD}$. 
We believe that a $c/l$ term will show up once quantum fluctuations are
incorporated as was the case in \cite{Luscher}
Unlike the MQCD result,  there is no constant term in the expression derived
from the supergravity setup \cite{GrOl}, but this may follow from a
different subtraction.     

In this work on precision ``measurements" of the quark potential from MQCD, 
we found two types of corrections to the linear potential 
between external (very heavy) quark and anti--quark in $SU(2)$,
  ${\cal N}=1$,  SYM. 
The first is a constant correction proportional to 
$R$ (the radius of the 11-th dimension) and the second is essentially
exponential in $-\Lambda_{QCD} \: l$. 
We considered two configurations -- one with both quarks on the same 
NS5 brane, and the other with the quarks on the two opposite NS5 branes. 
Although these configurations are supposed to describe the same field theory, 
we found the corrections are different.
We looked at theories where supersymmetry is softly broken and found that the behavior is qualitatively the same - a linear potential with a constant and exponential corrections. 
For more general $SU(N)$ (S)YM one can write analogous equations 
to (\ref{lofe},\ref{eofe}).
However, as is clear from (\ref{lofe}), for $N\neq 2$ the expressions will be
much more cumbersome. 

\subsection{Acknowledgments}
We are happy to thank Ofer Aharony, Oded Kenneth and Shimon Yankielowicz
 for very valuable discussions.

\section{Appendix - The explicit computation of the asymptotics of $l$ and $E$}
\subsection{Asymptotics of $l$}
The computation involves extracting the asymptotic behavior of $l(\e)$,  
and substituting its inverse in the asymptotic expression of $E(\e)$.
The leading behavior of $l(\e)$ reads
\begin{equation}
\label{lofe3}
l = \left(-\sqrt{2} \log \e + 5 \sqrt{2} \log 2 - {3 \over {4 \sqrt{2}}} \e \log \e + O(\e) \right) R
\end{equation}

In order to prove that, we first show that
\begin{equation}
\label{I1}
I_1(\e) \equiv \int_0^\infty {{dy} \over {\sqrt{y (y+\e) (2+y+\e)}}} = -{1 \over \sqrt{2}} \log \e + {{5 \log 2} \over \sqrt{2}} + {1 \over {8 \sqrt{2}}} \e  \log \e + O(\e)
\end{equation}

This behavior can be calculated in a few steps. First we look at an integral with the same divergence for $ \e \approx 0$ as $I_1(\e)$, for which we can find a primitive function and therefore evaluate its value as a series in $\e$ :
\begin{equation}
I_2(\e) \equiv \int_0^\infty {{\sqrt{2} dy} \over \sqrt{y (y+\e)(y+2)^2}} = -{1 \over \sqrt{2}} \log \e + {{3 \log 2} \over \sqrt{2}} - {1 \over {4 \sqrt{2}}} \: \e \log \e + O(\e)
\end{equation} 

All that remains is to evaluate the contribution of the difference $I_1 - I_2$. As both integrals have the same divergence for $\e \approx 0$ the difference is continuous for $\e = 0$ and its leading contribution is given by setting $\e = 0$ : 
\begin{equation}
\Delta I(0) \equiv I_1(0) - I_2(0) = \int_0^\infty dy {{\sqrt{2+y} -\sqrt{2}} \over {y (2+y)}} = \sqrt{2} \log 2
\end{equation}

The  $ \e \, \log \e $ term of the difference is found by looking at the derivative with respect to $\e$ :
\begin{equation} 
{d \over {d \e}} \Delta I(\e) = \int_0^\infty dy {{ -\e (4 + 2 y + \sqrt{2} \sqrt{2 + \e + y}) - (2+y) ( 2 + 2 y - \sqrt{2} \sqrt{2 + \e+y})} \over {2 \sqrt{y} (2+y) (2+\e+y)^{3/2} (\e+y)^{3/2}}}
\end{equation}

Again, we shall look for an integral with the same divergence for $\e \approx 0$, and having a primitive function. This time we have:
\begin{equation}
\int_0^\infty dy {{-(3y+\e)} \over {\sqrt{2 y} (\e+y)^{3/2} (2+y)^3}} = {3 \over {8 \sqrt{2}}} \log \e + O(1)
\end{equation}

(As before, checking that the difference of the integrals does not diverge when we set $\e = 0$ provides a proof that we found the correct diverging term). 
Integrating the result we got with respect to $\e$, we find that 
the difference's contribution is ${3 \over {8 \sqrt{2}}} \, \e  \log \e $,
and show that $I_1(\e)$ is as in (\ref{I1}). 
Inserting this expression in (\ref{lofe}) 
\begin{equation}
l = 2R\sqrt{(1+\e)} \int_0^\infty {dy \over \sqrt{y(y+\e)(y+2+\e)}} 
\end{equation}
we finally find (\ref{lofe3}). 
All that is left is to invert this expression, namely 
\begin{eqnarray}
\log (\e / 32) & = & - {l \over {\sqrt{2} R}} \left( 1 - 12 e^{-l/\sqrt{2}R}\right) + O(e^{-l/\sqrt{2}R}) \\
\label {eofl}
\e / 32 & = & e^{-l/\sqrt{2}R} + {12 \over \sqrt{2}} (l/R) e^{-2l/\sqrt{2}R} + O(e^{-2l/\sqrt{2}R})
\end{eqnarray}
\subsection{Asymptotics of $E$}
The expression for $E(l(\e),\e)$ was given in  (\ref{eofe})
\begin {equation}
E  =  2 \sqrt {2 \zeta} \sqrt{1+\e} \:l + 4 \sqrt 
{2 \zeta} R \: J(\e) - 4 \sqrt {2 \zeta} R \int_1^{1+\e} {{dz \sqrt{z}} \over \sqrt{z^2 - 1}}
\end{equation}
It is straightforward to convert the first term  into 
\begin{equation}
2 \sqrt {2 \zeta} \sqrt{1+\e} \:l = \left( 1 + {1 \over 2}\e + O(\e^2)\right) 2 \sqrt {2 \zeta} \:l 
\end{equation}

For the third term we use  $ z \approx 1 $ within the whole integration range 
so that  we can express the integrand as a series in $z-1$:

\begin{eqnarray}
\int_1^{1+\e} {{dz \sqrt{z}} \over {\sqrt{z^2 - 1}}} & = & \int_1^{1+\e} dz \left( {1 \over {2 \sqrt{z-1}}} + {\sqrt{z-1} \over {4 \sqrt{2}}} + O((z-1)^{3/2})\right) \nonumber \\
      & = & \sqrt{2} \cdot \sqrt{\e} + {1 \over {6 \sqrt{2}}} \e ^{3/2} + O(\e^{5/2})
\end{eqnarray}

The second term of (\ref{eofe}) has a finite negative contribution 
for $ \e = 0 $. We define 
\footnote{$\kappa$ can also be given a closed form in terms of 
elliptic functions.} 
\begin{equation}
\kappa \equiv -J(0) \approx 1.63029
\end{equation}

and so this finite contribution is
\begin{equation}
4 \sqrt {2 \zeta} \, R \: J(0) = - 2\kappa \cdot 2\sqrt{2 \zeta} \, R
\end{equation}

To find the sub--leading order behavior in $ \e $ we write the integral $J(\e)$ as a sum of three integrals 
\begin{eqnarray}
J(\e) & \equiv & J_3(\e) + J_1(\e) - J_2(\e) \nonumber \\
      & \equiv & \int_1^\infty dy {{\sqrt{y} - \sqrt{y+1+\e}} \over \sqrt{(y+\e)(2+y+\e)}} + \int_0^1 dy {\sqrt{y} \over \sqrt{(y+\e)(2+y+\e)}} \\
      &        &  - \int_0^1 dy {\sqrt{y+1+\e}  \over \sqrt{(y+\e)(2+y+\e)}} \nonumber
\end{eqnarray}

$J_3(\e) $ is an analytic function at $\e=0$, so it is possible to write it as a series in $\e$  
\begin{eqnarray}
J_3(\e) & = & \int_1^\infty dy  {{\sqrt{y} - \sqrt{y+1}} \over \sqrt{y(y+2)}} \nonumber \\
    &    &   - \int_1^\infty dy \left({1 \over {2 \sqrt{1+y} \sqrt{y(2+y)}}} + {{(1+y)(\sqrt{y} - \sqrt{y+1})} \over {y(2+y) \sqrt{y(2+y)}}} \right) \: \e+ O(\e^2)  \\
        & = & C_3 + \e \left( \sqrt{2 \over 3} - {1 \over \sqrt{3}} + {\log {2} \over {2 \sqrt{2}}} - {1 \over \sqrt{2}} \log {(2 + \sqrt{6})} \right) + O(\e^2) \nonumber
\end{eqnarray} 

The behavior of $J_1(\e)$ and $J_2(\e)$ can be found in a similar way 
to the one we used to find $l$: we find a function with the same divergence, 
subtract, differentiate with respect to $\e$ and repeat the process . 
Using this method we find: 
\begin{eqnarray}
J_1(\e) & = &C_1 + \e \left({1 \over \sqrt{3}} + { \log{(2+\sqrt{6})} \over \sqrt{2}} - {3 \over \sqrt{2}} \log 2 \right) + {1 \over {2 \sqrt{2}}} \: \e \, \log \e \nonumber \\
        &   &  - {1 \over {32 \sqrt{2}}} \: \e^2 \, \log \e + O(\e^2) \\
J_2(\e) & = & C_2 - \sqrt{2} \cdot \sqrt{\e} + \sqrt {2 \over 3} \e - {1 \over {6 \sqrt{2}}} \e ^{3/2} + O(\e^2)
\end{eqnarray} 

Summing up all the contributions to the energy we have 
indeed shown (\ref{Eofe2})  
\[ 
E = 2 \sqrt{2 \zeta} \: l \left( 1 + {1 \over 2} \e - {1 \over 8} \e^2 + O(\e^3) + 2 (R/l) (- \kappa + {1 \over {2 \sqrt{2}}} \: \e \, \log {\e \over 32} - {1 \over {32 \sqrt{2}}} \: \e^2 \, \log {\e \over 32} + O(\e^2)) \right)
\]

\end{document}